\documentclass[aps,prl,showpacs,twocolumn,superscriptaddress]{revtex4}
\usepackage{graphicx}

\begin{document}
\bibliographystyle{prsty}

\title{Modeling Two Dimensional Magnetic Domain Patterns}
\date{\today}

\author{J.R. Iglesias}
\affiliation{Instituto de F\'{\i}sica, Universidade Federal do Rio
Grande do Sul, C.P. 15051, 91501-970 Porto Alegre, Brazil}
\affiliation{Laboratoire de Physique des Solides, Universit\'{e}
Paris-Sud, B\^atiment 510, 91405 Orsay, France}

\author{S. Gon\c{c}alves}
\affiliation{Instituto de F\'{\i}sica, Universidade Federal do Rio
Grande do Sul, C.P. 15051, 91501-970 Porto Alegre, Brazil}

\author{O. A. Nagel}
\affiliation{Departamento de F\'{\i}sica, Universidad Nacional del
Sur, Av. Al\'{e}m 1253, (8000) Bah\'{\i}a Blanca, Argentina}

\author{Miguel Kiwi}

\affiliation{Facultad de F\'{\i}sica, Pontificia Universidad
Cat{\'{o}}lica de Chile, Casilla  306, Santiago, Chile 6904411 }

\begin{abstract}

Two-dimensional magnetic garnets exhibit complex and fascinating
magnetic domain structures, like stripes, labyrinths, cells and mixed
states of stripes and cells. These patterns do change in a reversible
way when the intensity of an externally applied magnetic field is
varied. The main objective of this contribution is to present the
results of a model that yields a rich pattern structure that closely
resembles what is observed experimentally. Our model is a generalized
two-dimensional Ising-like spin-one Hamiltonian with long-range
interactions, which also incorporates anisotropy and Zeeman terms.
The model is studied numerically, by means of Monte Carlo
simulations. Changing the model parameters stripes, labyrinth and/or
cellular domain structures are generated. For a variety of cases we
display the patterns, determine the average size of the domains, the
ordering transition temperature, specific heat, magnetic
susceptibility and hysteresis cycle. Finally, we examine the
reversibility of the pattern evolution under variations of the
applied magnetic field. The results we obtain are in good qualitative
agreement with experiment.

\end{abstract}


\maketitle

\input epsf

\section{Introduction}
\label{sec:intro}

Magnetic domains in magnetic garnet films are interesting in
themselves because of the rich and complex pattern formation they
display; stripes, cells, labyrinths, ordered hexagonal lattices,
disordered froth, pentagonal structures along the front between the
ordered lattice and the disordered froth, mixed states of stripes and
cells, are some of the features observed
experimentally~\cite{eschen,wea-riv,bab1,bab2}. In addition, these
patterns are quite universal, since they are similar to those
observed in soap froth and in the annealing process of
polycrystalline materials~\cite{kurtz,glaz}. Moreover, the
transitions are driven by stress induced by an applied magnetic bias
field H, a handy control parameter, which plays a role analogous to
temperature and/or pressure in the melting transition of solids.

\begin{figure}[h]
\centering
\resizebox{\columnwidth}{!}{\includegraphics{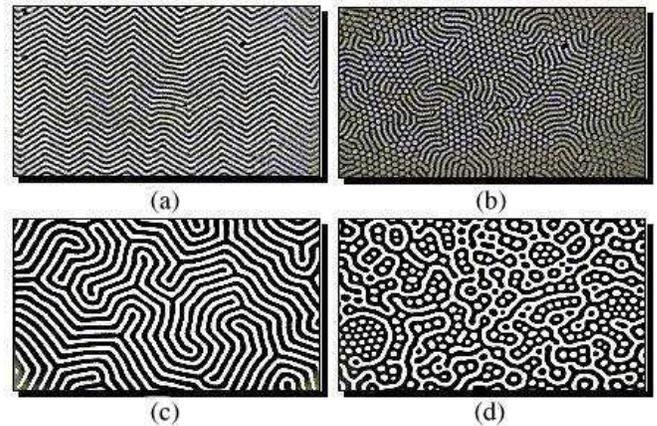}}
\caption{Stable domain structures observed experimentally by
  Albuquerque~\cite{thesis} in a magnetic  garnet. Obtained by
  varying the combination of perpendicular to the film and in plane
  applied magnetic fields.}
\label{fig:molho1}
\end{figure}

Originally the structure of magnetic domains in ultra-thin magnetic
garnet films attracted attention due to their potential application
as magnetic bubble memories~\cite{eschen}. In fact, size and thermal
stability of the domains are crucial to optimize recording
performance. Due to the interesting physics related to these
technological problems, significant  basic research efforts have
been carried out lately ~\cite {yeomans,sampaio,molho}. The most
extensively investigated systems are single crystal ferromagnetic
thin films of the FeYBiGdGaO garnet type. They are 5 to 7~$\mu $m
thick and exhibit a considerable anisotropy along the direction
perpendicular to the film, so that the magnetization remains along
this axis except within the Bloch walls that separate opposite
direction domains. These Bloch walls are of the order of 0.1~$\mu $m
thick. Different types of stable and metastable domain structures are
generated in such materials~\cite {bab2,yeomans,molho,thesis,weaire},
depending on the transient magnetic fields applied during the
thermo-magnetic cycling of the sample. Eye-appealing examples of the
most common structures observed experimentally were obtained by
Albuquerque and are found in reference~\cite{thesis}. The most
prototypical examples, displayed in Figs.~\ref{fig:molho1}, can be
grouped into several categories: parallel stripes
[Fig.~\ref{fig:molho1}(a)], mixed states of labyrinths and bubbles
[Fig.~\ref{fig:molho1}(b)], pure labyrinths
[Fig.~\ref{fig:molho1}(c)], and cell structures with a strong
labyrinth component, as illustrated in Fig.~\ref{fig:molho1}(d)

Often the bubble structure looks like soap froth, which originated
the denomination ``magnetic foam''. These magnetic foams are
generated by cooling the sample, initially at a temperature $T$
larger than the critical one $T_c$, to $T<T_c$ in the presence of
a weak magnetic field perpendicular to the plane of the film. The
magnetic bubble patterns result from a balance between the
demagnetizing energy (like long range dipolar interactions) and
the energy required to create Bloch walls.  This balance
originates an equilibrium domain size, which evolves as a function
of the strength of the applied magnetic field. When this external
field is increased, the bubble structure coarsens and large
bubbles, surrounded by smaller ones are observed, while the total
number of bubbles decreases.

Another intriguing feature of these foams is the memory effect. It
consists in a reversible change of the foam topology under a cyclic
variation of the applied magnetic field. For example, the bubble
configuration observed under an applied magnetic field evolves into a
labyrinth pattern at zero field, but the system returns to the previous
macroscopic configuration when the magnetic field is restored. It is
the stability of the threefold vertices that seems to be responsible
for this interesting memory effect~\cite{wea-riv,bab2}.

Over the years significant effort has been devoted to the understanding of
this rich and complex behavior~\cite{eschen,wea-riv}. For the particular
case of magnetic froth Weaire {\it et al.}~\cite{weaire} proposed a
description based on an analogy with two-dimensional soap froth, including
surface and compressibility energy terms. Sampaio {\it et al.}~\cite{sampaio}
proposed a model Hamiltonian to describe the formation of magnetic domains
in strongly anisotropic magnetic thin films, on the basis of a ferromagnetic
Ising Hamiltonian, including long-range (antiferromagnetic) dipole-dipole
interactions and an external magnetic field.

In this contribution we put forward an alternative model, addressing
the problem of the magnetic domain pattern formation by means of a
generalized spin Hamiltonian. More precisely, we consider a
three-state spin one Ising Hamiltonian, with competing long-range
interactions, and we also include the anisotropy energy and an
external field.

This paper is organized as follows: after this Introduction, in
Sec.~\ref{sec:model} we present the model and the simulation
technique. In Sec.~\ref{sec:results} the results reached by
implementing our model are presented. The paper is closed in
Sec.~\ref{sec:concl}, where we provide a summary and draw conclusions.

\section{Hamiltonian and Simulation}
\label{sec:model}

\subsection{\protect\bigskip Hamiltonian}
\label{subsec:H}

We model the system by a generalized spin one Ising-like Hamiltonian
on a square lattice. The interaction between nearest neighboring
spins is ferromagnetic and long-range competing dipolar interactions,
between the $n$ nearest neighbor sites ($n > 1$), are incorporated. We
also include the magnetic field and the anisotropy. Analytically our
model Hamiltonian is

\begin{eqnarray}
\label{Eq:model} H & = & - \frac 12 {\sum_{i,j}}^{\prime} J_{ij}
S_{i}^{(z)} S_{j}^{(z)} - \sum_{i} [A (S_{i}^{(z)})^{2} \nonumber\\
& & + {\text H}_\ell S_{i}^{(z)} + {\text H}_{t}\{ 1-
(S_{i}^{(z)})^{2}\}]  \; ,
\end{eqnarray}

\noindent where the parameters $J_{ij}$ are the exchange coupling
constants between the spin on the $i$-site and its first $n$ nearest
neighboring spins, located on sites $j$. Thus, $J_{ij}>0$
($J_{ij}<0$) imply ferromagnetic (antiferromagnetic) coupling. The
anisotropy constant is defined such that $A>0$ favors spin
orientation perpendicular to the plane of the sample.
${\text{H}}_{\ell }$ and ${\text{H}}_{t}$ are the intensity of the
longitudinal and transverse external magnetic fields, respectively,
and the $i$ and $j$ summations are performed over $N(N-1)$ lattice
sites, since the primed summation implies that the self-interaction
terms $i=j$ are excluded.  The transverse magnetic field and the
anisotropy play a similar role, but while $A$ may be considered as a
microscopic characteristic of the system, ${\text{H}}_{\ell }$ and
${\text{H}}_{t}$ are externally controlled parameters.

If one chooses for the coupling constants $J_{ij}$ the dipolar
interaction a square-angled labyrinth is generated, as in
reference~\cite{sampaio}. Instead, we adopt a long-range oscillatory
potential of the RKKY-type. This way, and although the system is not
metallic, use of an RKKY-like expression provides a set of oscillating
coupling constants that depends on a single parameter $k$. The
analytical expression we propose for $J_{ij}$ is given by

\begin{equation}
J_{ij} = J_0 \frac {\cos(k r_{ij})} {(k r_{ij})^3} \; ,
\label{eq:couplings}
\end{equation}

\noindent where $r_{ij}=|\vec{r}_{i}-\vec{r}_{j}|$ is the distance between
sites $i$ and $j$. In the numerical calculations we consider
interactions up to the seven-th nearest neighbor and for convenience
we adopt as our unit of energy, here and throughout, the first
neighbor exchange constant $\Gamma$, defined by

\begin{equation}
\Gamma =J_{0}\frac{\cos (ka)}{  (ka)^{3}}=1 \; ,
\label{eq:gamma}
\end{equation}

\noindent where $a$ is the lattice parameter. This type of coupling
guarantees that the interactions between nearest neighbors is always
ferromagnetic and of the same strength, while the magnitude and sign
of the coupling with the rest of the neighbors depends on $k$.

Thus, our Hamiltonian differs from previous proposals, like the one
studied by Sampaio {\it et al.}~\cite{sampaio} in at least three
important aspects. We adopt: i)~spin 1, instead of spin spin $\frac{1
}{2}$, so that the domain walls may adopt finite width, since spin
$\frac{1 }{2}$ implies abrupt transitions from one domain to its
neighboring one; ii)~a long-range interaction that is not purely
antiferromagnetic and decays as $ r_{ij}^{-3}$; and
iii)~magnetic anisotropy is incorporated.  As it will become clear
hereafter, the inclusion of these three effects leads to significant
qualitative and quantitative differences in the results, and allows
for a better description of experimental observations.

\subsection{Simulation procedure}
\label{subsec:sim&procedure}

All of our calculations were performed on a $N\times N$ lattice, with
$N=200$, using a conventional Monte Carlo (MC) procedure to update the
lattice spins at various temperature, $T$, and applied field, H,
values. One Monte Carlo step (MCS) consists in the successive updating
of ${N^{2}}$ spins chosen at random, following the Metropolis
procedure.  Temperatures and energies are measured in units of $\Gamma
=1$, and $k$ in units of the inverse lattice parameter $a^{-1}$.

Two different processes were investigated: quenching and relaxation.
{(a)}~{\it Quenching}: In order to quench the system the lattice
temperature $T$ is raised to infinity (all spin-states generated at
random) and at the first Monte Carlo step $T$ is reduced to a very
low value ($T=0.005$), after which the system is allowed to evolve
for 1000~MCS.  This way the system is prevented from exploring most
of the available energy landscape and quite probably it is quenched
into a metastable minimum energy state, different from the lowest
minimum.  {(b)}~{\it Relaxation}: On the other hand, thermal
relaxation is achieved by slowly varying the temperature from a large
initial value ($T=5$) to a final temperature $T=0$, in intervals of $
\Delta T=0.05$.  To be certain that thermal relaxation is indeed
achieved we run 1000~MCS after every temperature step; therefore, for
a typical thermal relaxation run a total of $\approx 10^{5}$ MCS are
required, which is equivalent to $4 \times 10^{9}$ individual spin
flips.

During these processes, and as will be described in detail in
Sec.~\ref{sec:results}, several thermodynamic properties and
parameters of the system are monitored: the energy $E$, the
magnetization $M$, and the average domain size $\Lambda $ are
evaluated during the {\it quenching} process; along the {\it
relaxation} process the specific heat $C$ and the magnetic
susceptibility $\chi$ were are also monitored, as a function of $T$.
In the latter case we perform an average over 20 different replicas
of the system in order to decrease the thermal fluctuations.
The domain size $\Lambda$ was evaluated by means of a
Hoshen-Kopelman~\cite{kopelman1,kopelman2} type algorithm. This
modified algorithm computes the cluster size identifying all
clusters, calculating their surface and degeneracy, and then
evaluating the average cluster size. The specific heat $C$ is calculated
by means of the energy fluctuations~\cite{NewBar}, and is given by

\begin{equation}
C = \frac {\langle E^{2}\rangle -\langle E\rangle ^{2}} {(k_B T)^{2}} \; .
\label{eq:C}
\end{equation}

Finally, the spin susceptibility $\chi$ at zero field is obtained
from the magnetization fluctuations, and given by

\begin{equation}
\chi = \frac {\langle M^{2}\rangle -\langle M\rangle ^{2}} {k_B T}\; .
\label{eq:chi}
\end{equation}

\section{Results}
\label{sec:results}

\subsection{Model parameters}
\label{subsec:parameters}

\begin{figure}[h]
\centering
\resizebox{\columnwidth}{!}{\includegraphics{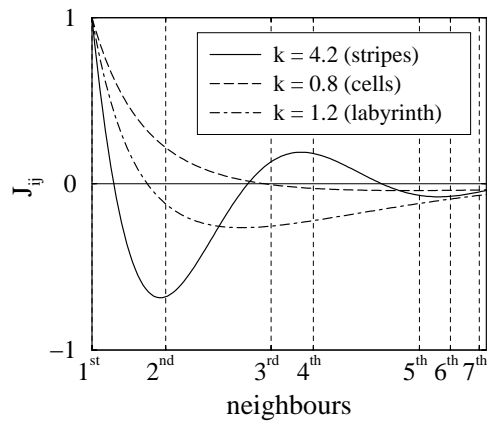}}
\caption{Long range  RKKY-like dipolar potentials used in
  the simulations.}
\label{fig:rkky}
\end{figure}

The first part of our work was devoted to search for an adequate set
of parameters, that generates magnetic domain patterns like the ones
observed in magnetic foam experiments~\cite{molho,thesis}, and
displayed in Figs.~\ref{fig:molho1}. The
procedure we employ is the following: we adopt a value of $k$, that
in turn determines the values of the exchange parameters $J_{ij}$ up
to the seven-th nearest neighbor, without including anisotropy
($A=0$) and for zero applied magnetic field (H~$=0$). Next, the
system is quenched from $T\longrightarrow \infty$ to a very low $T$
value ($T=0.005$). At the end of the run we keep or discard the
choice of parameters by visual inspection.  After selecting some
interesting values of $k$ we varied the anisotropy $A$.  The $k$
values we finally chose, following the above outlined scheme, yield the
long range exchange functions illustrated in Fig.~\ref{fig:rkky}.

\begin{figure}[h]
\centering
\resizebox{\columnwidth}{!}{\includegraphics{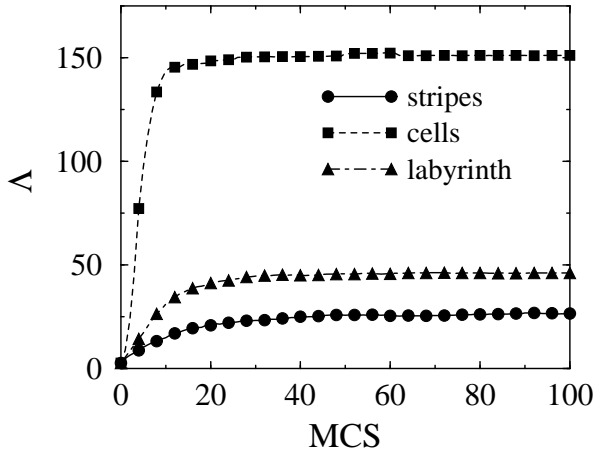}}
\caption{Average domain size $\Lambda$ versus number of
Monte-Carlo
  steps (MCS).}
\label{fig:SvsMCS}
\end{figure}

All the configurations are obtained after 1000~MCS, as explained in
Sec.~\ref{subsec:sim&procedure}.  This number of steps is far larger
than what is required to reach steady state. To stress this fact we
display in Fig.~\ref{fig:SvsMCS} the domain size $\Lambda$ versus the
first 100 MCS is displayed. It is clearly noticed that after the
first 30 MCS the system reaches steady state values. It is also
apparent that stripes and labyrinths are significantly smaller than
the average cell dimension.  Along with $\Lambda$, the total energy
$E$ and the magnetization $M$ were also computed, but are not
displayed graphically here as their time variation is similar to that
of $\Lambda$.  Thus, all in all it is evident from the behavior of
these quantities, and from visual inspection of Fig.~\ref{fig:SvsMCS},
that the system reaches steady state (metastable) equilibrium after
relatively few MCS.

\begin{figure}[h]
\centering
\resizebox{\columnwidth}{!}{\includegraphics{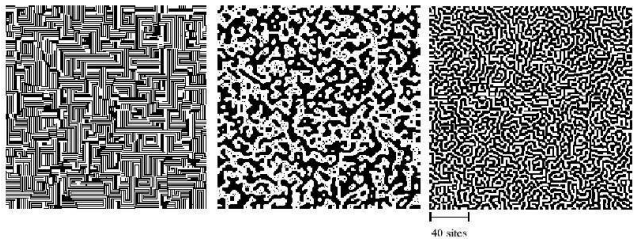}}
\caption{Snapshots of the three types of domain structure obtained
  through simulations, after quenching. The stripe patterns~(left)
  correspond to the parameters $k=4.2$ and $A=-0.5$~. The
  cells~(center) to $k=0.8$ and $A=-0.1$~. Finally, the
  labyrinths~(right) to $k=1.2$ and $A=-0.8$~.}
\label{fig:patterns}
\end{figure}

The domain structures generated, after quenching, give rise to
three different patterns: stripes, labyrinths and cells.  In all
cases negative values of the anisotropy $A<0$ were adopted, to
favor the presence of domain walls ($S=0$ sites). These sites
emerge preferentially between opposite magnetization domains,
since they lower the energy cost of growing a domain wall.
Fig.~\ref{fig:patterns} exhibits snapshots of the three types of
domain structures obtained after quenching. They correspond to the
following sets of parameters: i)~$k=4.2$ and $A=-0.5$ (stripes);
ii)~$k=0.8$ and $A=-0.1$ (cells); and finally iii)~$k=1.2$ and
$A=-0.8$ (labyrinths).

As stated in Sec.~\ref{sec:model} these patterns emerge from the
competition between the ferromagnetic first neighbor exchange
interaction $\Gamma>0$, and the long-range oscillating
interactions given by Eq.~\ref{eq:couplings}.  For the stripe
structure ($k=4.2$, Fig.~\ref{fig:patterns}a) the second neighbor
exchange constant is antiferromagnetic and comparable in magnitude
with the nearest neighbor one, while the third and fourth are
ferromagnetic, leading to the formation of stripes of alternating
spin orientation. This way, two frustrated couplings between
nearest neighbors are compensated by four pairs of satisfied
second neighbor interactions. The labyrinth case ($k=1.2$),
illustrated in Fig.~\ref{fig:patterns}c, is also a consequence of
competing interactions between the ferromagnetic first neighbor
coupling and the rest, all of which are now negative and smaller
in magnitude. In this case stripe domains, one lattice parameter
wide, do not optimize the energy thus yielding a domain width
larger than in the previous case. Finally, when cells are
generated (see Fig.~\ref{fig:patterns}b), both the first and
second neighbor interactions are ferromagnetic, and all the rest
are antiferromagnetic and of much smaller magnitude. Consequently,
the system approaches full ferromagnetism and a quenched cell
structure results from a frozen metastable spinodal decomposition.
Later we will see that this cell structure, as well as the typical
eddy structures of the labyrinth case, both originate in the
quenching process.

Before going on, it is important to stress the fact that each of
the configurations in Fig.~\ref{fig:patterns} originates solely
from different Hamiltonian parameters, since both $J_{ij}$ (which
is a function of $k$) and $A$ do vary. On the other hand, in the
experiments each domain pattern is a metastable configuration
driven by an external applied bias field H, as mentioned in the
introduction. The effects of H on the domain patterns and on other
physical properties will be addressed in
Secs.~\ref{subsec:hysteresis} and \ref{subsec:memory}.

\subsection{Characterization}
\label{subsec:characterization}

Our main objective is to contribute toward an understanding of the
physical mechanisms responsible for the many qualitatively different
configurations that two-dimensional magnetic systems do adopt. Thus,
in order to properly characterize and systematize our conclusions,
several tools and physical parameters are used to illustrate the
results we obtain.  Clearly, direct pictures are the most
straightforward form of visualization, and consequently we start
providing representative snapshots of some of the typical patterns
that the system adopts.

\begin{figure}[h]
\centering
\resizebox{\columnwidth}{!}{\includegraphics{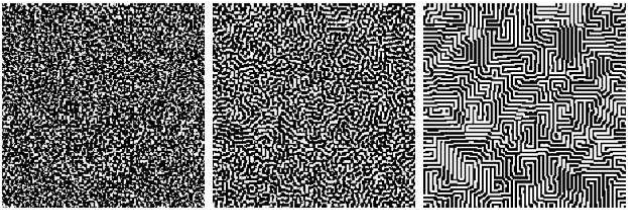}}
\caption{Snapshots that illustrate the results of simulating
  the relaxation process for the labyrinth case, starting from a large
  temperature $T\gg 1$: at $T=2$ a cell-like structure~(left) develops.
  The ordering process starts around $T=0.8$, as illustrated in~(center).
  Finally, an ordered stripe-like pattern is reached at $T=0.1$~(right).}
\label{fig:relax}
\end{figure}

To describe the equilibrium state we computed the specific heat
$C$, the magnetic susceptibility $\chi$, and the magnetization
$M$, as a function of temperature $T$.  In addition, in
Sec.~\ref{subsec:hysteresis} and \ref{subsec:memory} hysteretic
and memory effects, respectively, will be presented and discussed.

As already mentioned snapshots provide the most direct perception
and they allow to observe directly the evolution of the system.
The patterns displayed in Figs.~\ref{fig:relax} were obtained
during runs at various $T$ values, starting from a high initial
temperature, $k_B T>\Gamma$, as illustrated in
Fig.~\ref{fig:relax}a, where a cell-like structure develops. As
$T$ is lowered the ordering process is triggered when the critical
temperature $T=T_{c} $ is approached (Fig.~\ref{fig:relax}b
corresponds to $T=0.8$ and a labyrinth pattern does emerge).
Finally, the low temperature (Fig.~\ref{fig:relax}c illustrates
the $T=0.1$ result) the minimum energy configuration is attained.
As expected, the low $T$ stripe configuration reached in this way
exhibits a higher degree of order than the quenched one,
illustrated in Fig.~\ref{fig:patterns}c.  In other words, the eddy
patterns of the quenched labyrinth case are almost washed out by
thermal relaxation, evidencing that the latter correspond to a
metastable configuration. The experimental counterpart shows a
very similar behavior, as reported by Mino {\it et
al.}~\cite{mino1,mino2} for the (Ho$_{1.2}$Tb$_{0.6}$Bi$_{1.2}$)
Fe$_5$ O$_{12}$ compound, when the system was relaxed from the
quenched configuration under an oscillating magnetic field. This
provides support to the fact that our model constitutes a
reasonable choice.  Moreover, the cell case also corresponds to a
metastable pattern, obtained by freezing the spinodal
decomposition of a quasi-ferromagnet. This is confirmed by
relaxing the system, which leads to a perfect fully magnetized
($M=1$) ferromagnet.

\begin{figure}[h]
\centering
\resizebox{\columnwidth}{!}{\includegraphics{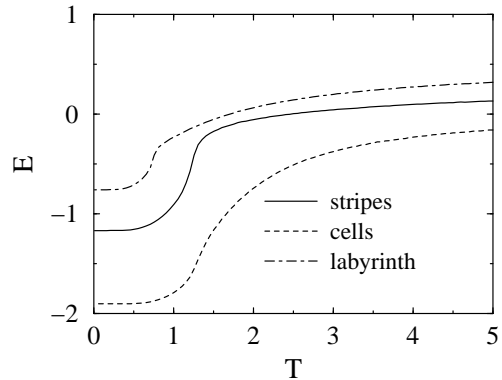}}
\caption{Energy  versus $T$ plots for the three types of
patterns.} \label{fig:EvsT}
\end{figure}

\begin{figure}[h]
\centering
\resizebox{\columnwidth}{!}{\includegraphics{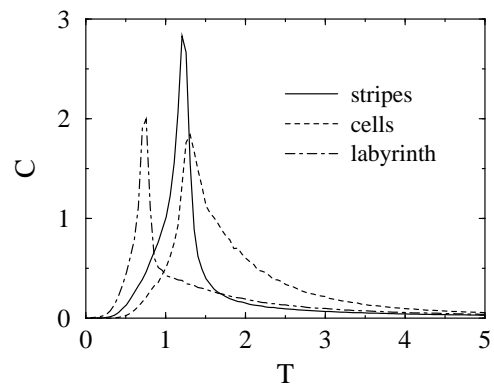}}
\caption{Specific heat versus $T$ plots for the patterns of
Fig.~\ref{fig:EvsT}.} \label{fig:CvsT}
\end{figure}

The MC thermal relaxation was carried out reducing the temperature
from $T=5$ to $T=0$ (we recall that energy units of $\Gamma=1$, as
defined in Eq.~\ref{eq:gamma}, are used throughout).  From the $E$
versus $T$ plot of Fig.~\ref{fig:EvsT} it is evident that a phase
transition takes place at the inflection point. It is apparent as
well that the energy sequence is
$E_{labyrinth}>E_{stripe}>E_{cell}$, as is verified by direct
inspection of Fig.~\ref{fig:EvsT}.  However, the most conclusive
evidence for the phase transition is the specific heat $C$, which
exhibits a characteristic peak (Fig.~\ref{fig:CvsT}) that
constitutes the signature of the transition.

We recall that this curve -and all the others obtained during the
relaxation processes- correspond to an average performed over 20
independent runs. This allows to appreciate the benefits of such a
procedure to attenuate the fluctuations inherent to computations on a
relatively small system, of only 200x200 sites, by comparison with
our preliminary results~\cite{iglesias} obtained with only one run.
It is observed that the ordering temperatures are of the same order
of magnitude as the nearest neighbor interaction $\Gamma$, which by
definition is the same for all the samples.  However, the critical
temperature $T_c$ is larger for the cell configuration, which is
mainly ferromagnetic, and lower for the stripe and labyrinth patterns.

\begin{figure}[h]
\centering
\resizebox{\columnwidth}{!}{\includegraphics{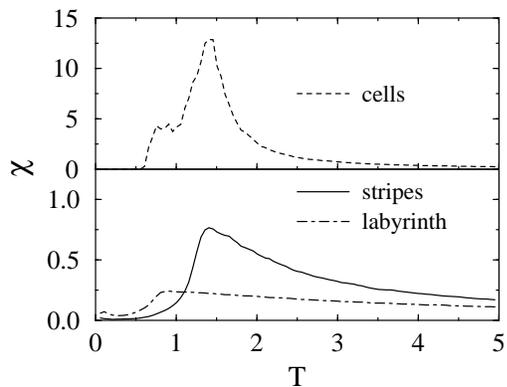}}
\caption{Susceptibility $\chi$ versus $T$ for the patterns of
Fig.~\ref{fig:EvsT}.} \label{fig:XvsT}
\end{figure}

For completeness we have also calculated the magnetization
fluctuations, {\it i.e.} the spin susceptibility $\chi$ at zero
field, and the corresponding results are displayed in
Fig.~\ref{fig:XvsT} for the three main structures we discuss. It is
quite apparent that stripes and labyrinths, which order mainly
antiferromagnetically, have a relatively low susceptibility, while
cells show a value of $\chi$ more than an order of magnitude larger,
consistent with their tendency to ferromagnetism.

\begin{figure}[h]
\centering
\resizebox{\columnwidth}{!}{\includegraphics{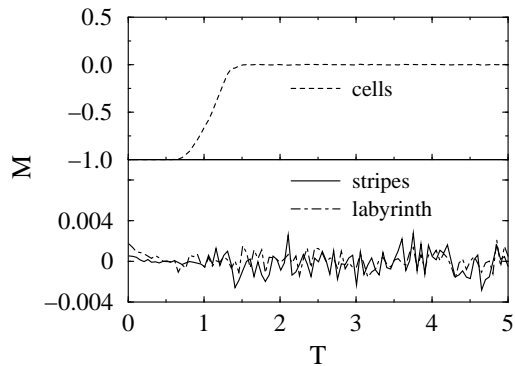}}
\caption{Magnetization M versus $T$ for the patterns of
Fig.~\ref{fig:EvsT}.} \label{fig:MvsT}
\end{figure}

Also, the thermal evolution of the magnetization, $M$ versus $T$,
portrayed in Fig.~\ref{fig:MvsT}, is an evidence that the stripe
and labyrinth structures are antiferromagnetic, as implied by the
low amplitude fluctuations close to $M=0$, with $\langle M \rangle
\approx 0$, in sharp contrast with the cell structure results
which exhibit a significant ferromagnetic magnetization. All the
above provides a picture that is self-consistent and agrees
qualitatively with experiment.

\subsection{Hysteresis cycle}
\label{subsec:hysteresis}

In order to characterize the system as completely as possible, we also
computed the magnetic hysteresis cycles by performing MC simulations
at temperatures $T<T_c\sim 1$, adopting the quenched state at zero
applied field (H~$=0$) as the initial configuration, and thereafter
increasing the magnetic field in increments of $\Delta $H~$=0.01$ (as
always, in units of $\Gamma$).  The magnitude of $M$ was computed
averaging over 1000 MCS for each value of the applied field intensity.
This process was performed until saturation of the magnetization was
achieved, after which the direction of the H field was reversed and
the full hysteresis cycle computed.  The results obtained are
consistent with the previous discussion.  The cell structure exhibits
a magnetic hysteresis typical of a soft ferromagnetic system: that is,
the system is easily saturated with a relatively low applied field
(H~$\approx 0.05$) and the remanent field is almost as large as the
saturation field.

\begin{figure}[h]
\centering
\resizebox{\columnwidth}{!}{\includegraphics{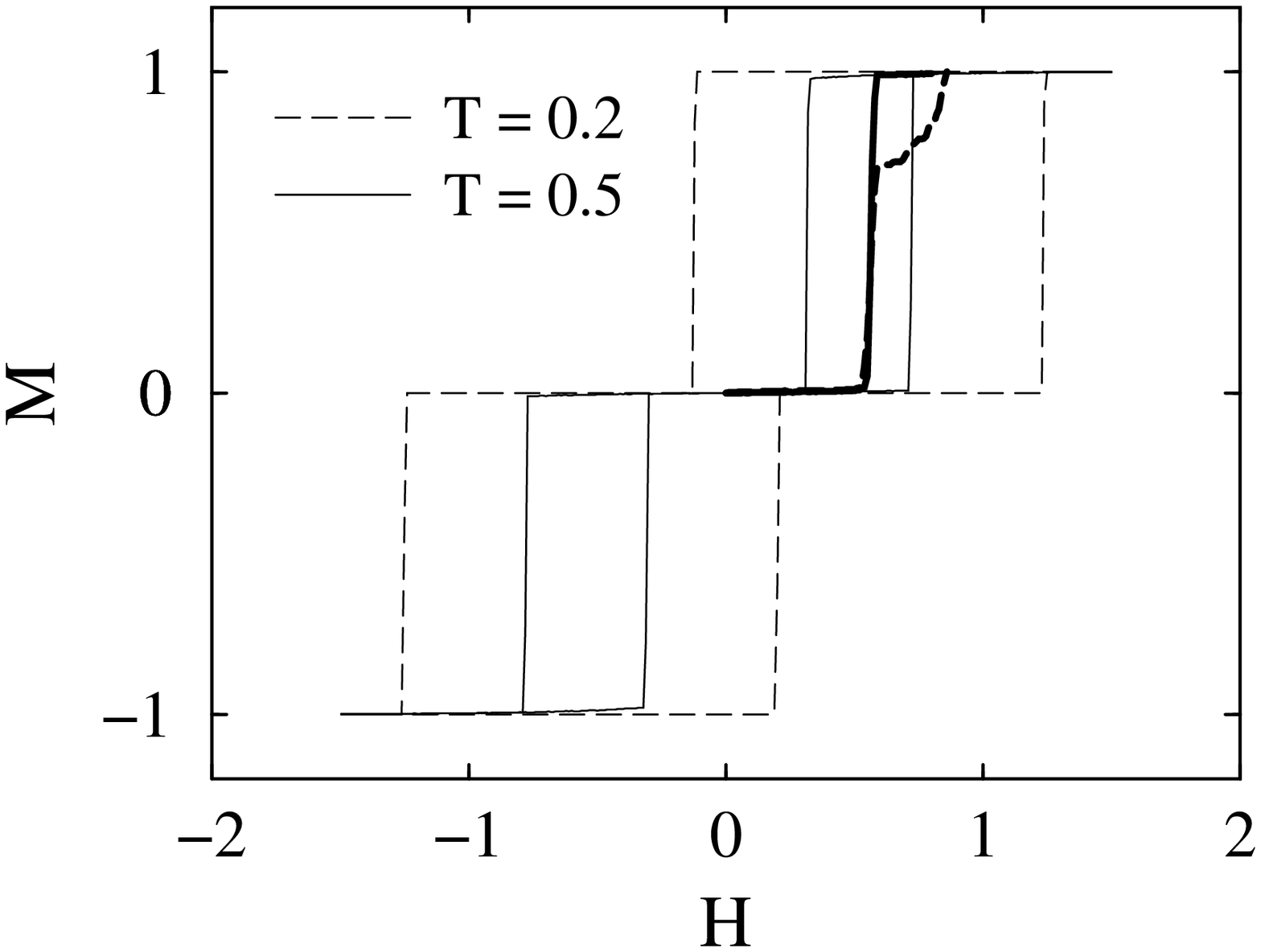}}
\centering
\resizebox{\columnwidth}{!}{\includegraphics{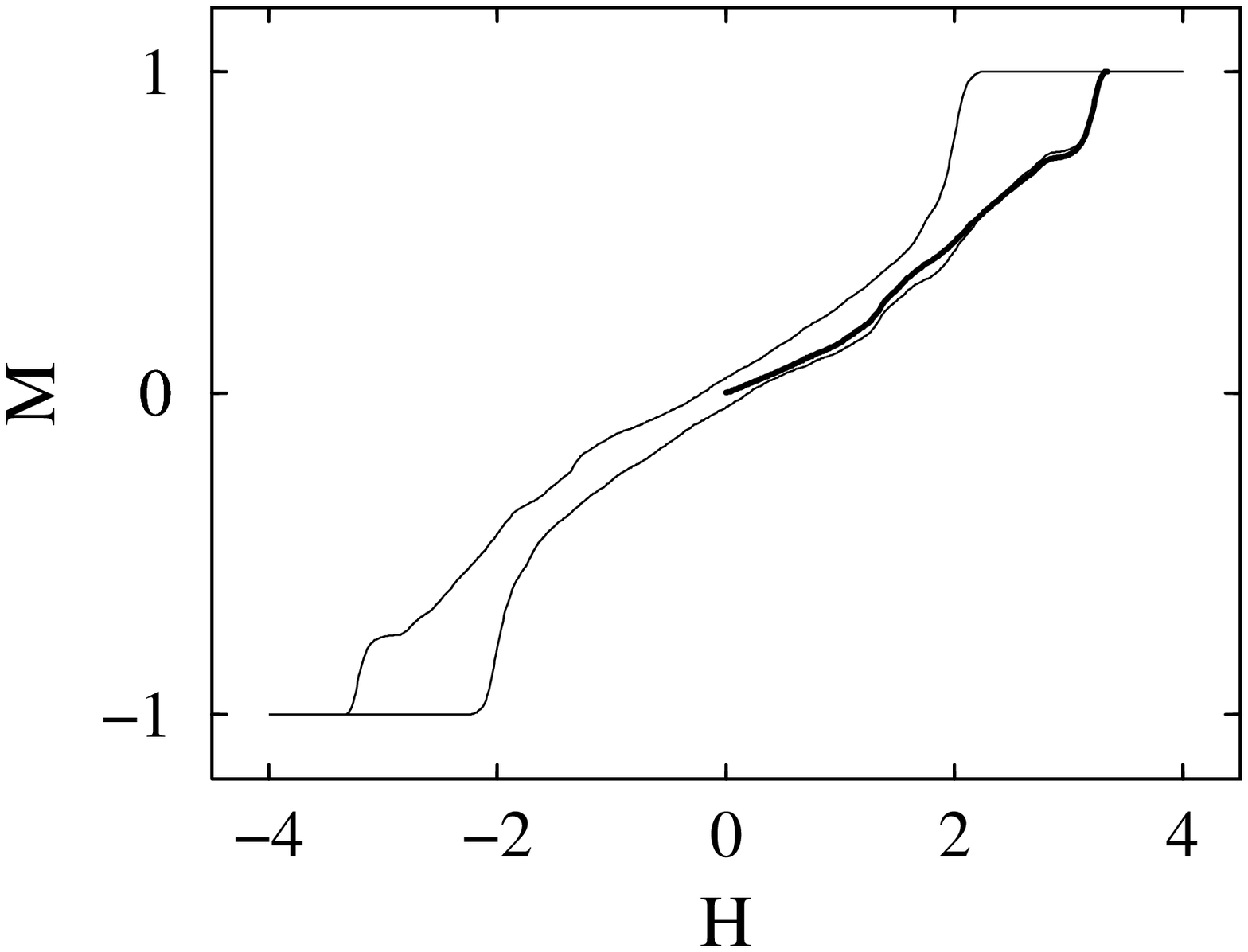}}
\caption{Hysteresis loops  (top)~stripes for $T=$~0.2 and 0.5;
  (bottom)~labyrinths for $T=0.05$~. The virgin curves are marked with
 thick lines.}
\label{fig:hist-loops}
\end{figure}

In sharp contrast the stripe structure, illustrated in
Fig.~\ref{fig:hist-loops}a, behaves quite differently. The
hysteresis loop consists of two nearly rectangular portions,
corresponding to positive and negative magnetizations $M$, shifted
away relative to one another. The width of these rectangular
portions shrinks considerably as the temperature is increased, but
the relative displacement remains nearly constant.  Thus, in spite
of the fact that the equilibrium configuration is
antiferromagnetic, at $T = 0.2$ the system remains saturated up to
H~$\approx 1.3$ implying that the energy landscape is bumpy and
with high energy barriers.  In order to verify the preceding
picture we also performed simulations at a higher temperature:
$T=0.5$. The corresponding hysteresis cycle is also shown in
Fig.~\ref{fig:hist-loops}a, and the dominant antiferromagnetic
character is again evident. The total magnetization has a sharp
transition from zero to fully saturated, without going through
intermediate values; as expected, the width of each of the
rectangular portions is significantly smaller than for $T=0.2$ and
saturation is obtained for applied fields H~$\approx 0.8$, {\it
i.e.}  not too different from the low temperature value.

On the contrary the labyrinth configuration, plotted in
Fig.~\ref{fig:hist-loops}b, where we represent the low temperature
$T=0.05$ cycle, is rather peculiar. It requires large H values to
reach saturation and the $M$ versus H cycle is almost reversible ({\it
  i.e.} with very low remanence).  Moreover, two different regimes are
clearly defined: for low fields the magnetization is essentially a
linear function of H, and it is in this region where one can expect to
obtain a memory effect. On the contrary, for large applied fields the
system saturates, but when H is decreased it returns to the linear
regime for the relatively large H~$\approx 2$ field value.

\subsection{Memory effect}
\label{subsec:memory}

\begin{figure}[h]
\centering
\resizebox{\columnwidth}{!}{\includegraphics{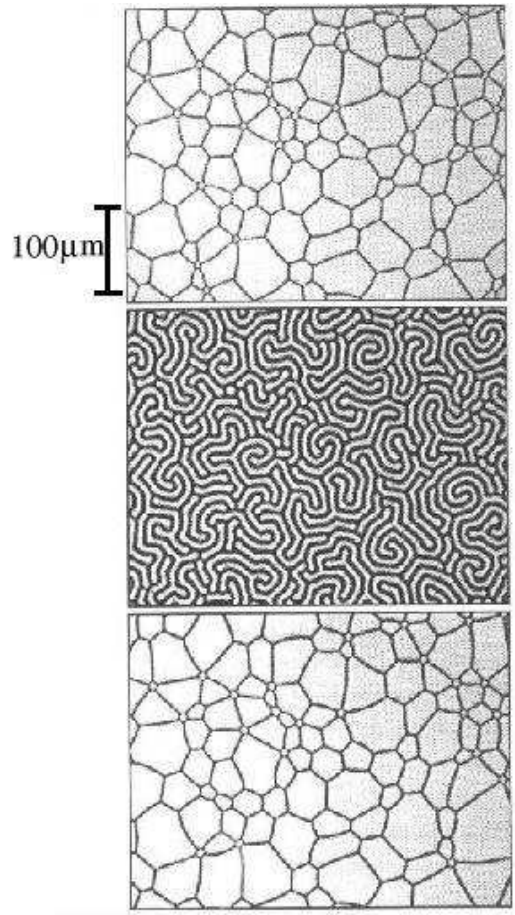}}
\caption{Memory effect observed by Albuquerque~\cite{thesis} under
  cycling of the applied magnetic field. The top figure corresponds to
  H = 78.9~Oe, the center one to H = 0 and the bottom figure again to
  H = 78.9~Oe. The close resemblance of the top and bottom figures is
  quite remarkable}
\label{fig:molho2}
\end{figure}

Among the results obtained by Albuquerque~\cite{thesis} perhaps
the most striking one is the visualization of the memory effect
illustrated in Figs.~\ref{fig:molho2}.  It consists in the
reversible change of the foam topology under a cyclic variation of
H.  The ``foam'' observed under an applied field H in the top
image of Fig.~\ref{fig:molho2} evolves into a labyrinth pattern at
zero field (middle figure), but the system recovers the initial
macroscopic configuration when H is restored to its original
value, as seen in the bottom image of Fig.~\ref{fig:molho2}.  It
has been conjectured that it is the stability of the threefold
vertices, that are so abundant in the upper and lower images
displayed in Fig.~\ref{fig:molho2}, which is responsible for this
interesting effect~\cite{wea-riv,bab2}.

\begin{figure}[h]
\centering
\resizebox{\columnwidth}{!}{\includegraphics{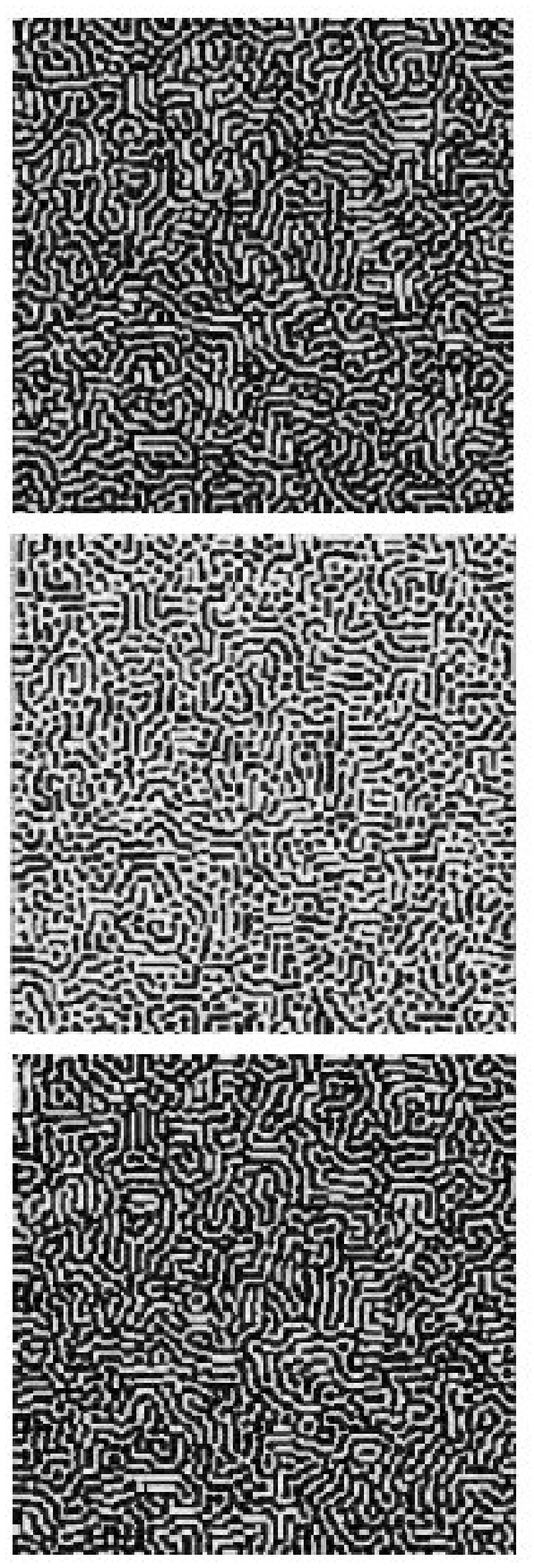}}
\caption{Simulation results of the memory effect of
  Fig.~\ref{fig:molho2} for the labyrinth case (H$_t=0.5$):
  (top)~H$_l=0$, (center) H$_l=1.5$, and (bottom) H$_l=0$~.}
\label{fig:memo}
\end{figure}

Unfortunately the memory effect is very difficult to simulate and
our efforts in this direction have met with only modest success.
In our quest we have investigated the labyrinth ($k=1.2$ and
$A=-0.8$) and cell ($k=0.8$ and $A=-0.1$) structures.  In both
cases we generated the initial structures by first quenching the
system and afterwards applying a transverse magnetic field
H$_t=0.5$ to favor the growth of a boundary (actually a ``skin'')
of $S=0$ spins that separates the $S=1$ from the $S=-1$ stripes.
It is important to stress that these Bloch walls (or ``skins'')
assemble as a very narrow domain wall (one or two lattice
parameters wide) but, in spite of their limited width, they are an
important element to generate the memory effect. As a matter of
fact, the effect fades away in the absence of these $S=0$ Bloch
walls. However, it should also be remarked that, due to the length
scale required to display our results in Figs.~\ref{fig:memo}, the
``skins'' are too narrow to be noticeable.

Following the quenching and the application of H$_t$ a
longitudinal field H$_\ell$, lower than the saturation value, is
also applied after which H is reduced to zero while performing
1000 or more MCS at every field value.  The labyrinth structure
thus obtained exhibits a sort of memory effect, as can be seen in
Figs.~\ref{fig:memo}a, \ref{fig:memo}b and \ref{fig:memo}c,
obtained for H$_\ell =0$, H$_\ell <1.7$ and H$_\ell =0$,
respectively (H$_t=0.5$ is kept constant during the cycle).  After
several cycles the structure does recover its overall original
appearance, thus yielding a genuine memory effect. However, there
are differences with the experimental results: both the
configuration we start from and the one we arrive at after cycling
differ, both as far as the shape of the labyrinths and the width
of the walls that separate them is concerned, with experimental
observation. On the other hand, in the cell structure case no
memory effect is observed, since the system is magnetically very
soft. Thus, repeated cycling just yields a fully polarized
lattice.

\section{Summary and Conclusion}
\label{sec:concl}

In this paper we model the magnetic relaxation and the formation of
magnetic patterns in ultrathin films with anisotropy perpendicular to
the film plane. To study this model the Monte-Carlo simulation
technique was implemented in combination with a generalized two
dimensional classical Ising-like Hamiltonian on a square lattice,
including long range interactions, and transverse H$_{t}$ and
longitudinal H$_\ell$ applied magnetic fields.  For the dipolar
interaction we adopted a long range oscillating coupling that extends
all the way to the seven-th nearest neighbor.

To present our results we start providing direct snapshots of the
patterns that develop in the simulations. In addition, and to
fully characterize the behavior of the system, we monitor the
evolution of the total energy, the magnetization, the average
domain size, the specific heat, the magnetic susceptibility and
the hysteresis cycle. Altogether they provide a physical
description which is qualitatively in agreement with experimental
results, yielding stripe, labyrinth and cell patterns.  The key
parameter that determines which type of pattern is generated is
the magnitude of $k$ (the argument of the RKKY part of the
potential given by Eq.~\ref{eq:couplings}) which modulates the
oscillations and the relative weight of the long range
interactions.  On the other hand, the specific heat versus
temperature plot implies the existence of an ordering temperature
$T_{c}$ below which pattern formation is observed after quenching
the samples.

Finally we address the striking memory effect observed by
Albuquerque~\cite{thesis}, which consists in the reversible change
of the foam topology under the cycling of the applied magnetic
field. While our model fails to yield the shape and size of the
domains observed experimentally, the patterns that evolve when the
cycling process is implemented do display the experimentally
established memory effect.

\acknowledgments

JRI, SG and MK acknowledge support from the Fundaci{\'o}n Andes, and
ON from the Centro Latino-Americano de F\'{\i}sica.  JRI, SG and ON
were supported by CNPq (Brazil) and MK by the {\it Fondo Nacional de
Investigaciones Cient\'{\i}ficas y Tecnol{\'o}gicas} (FONDECYT,
Chile) under grant \#8990005.


\end{document}